\documentclass[
 ,final            
 ,numberedheadings 
  ]
  {aipproc}

\layoutstyle{6x9}

\begin{document}

\title{Exclusive Meson Electroproduction: GPDs, Regge and Dispersion Relations}


\author{Gary R. Goldstein}{
  address={Tufts University, Department of Physics and Astronomy, Medford, MA 02155 USA.}
}

\author{Simonetta Liuti}{
  address={University of Virginia, Department of Physics,  Charlottesville, VA 22901, USA.}
}


\begin{abstract}
Exclusive pi0 electroproduction from nucleons at large photon virtuality can be described in terms of Generalized Parton Distributions, particularly the chiral odd subset. Chiral odd GPDs are related to transversity and can be accessed experimentally for various special choices of observables. This is accomplished by choosing C-parity odd and chiral odd combinations of t-channel exchange quantum numbers. These GPDs are calculated in a spectator model, and are constrained by boundary functions.  Alternatively, the production amplitudes correspond to C-odd Regge exchanges with final state interactions. The helicity structure of the virtual photoproduction amplitudes provides relations between the partonic description via GPDs and the hadronic, Regge description of C-odd processes, all via quark helicity flip. GPDs, in general, are analytic functions of energy variables. Their integrals over unobservable parton momenta thereby satisfy Dispersion Relations (DR). Using DR's could allow the real part of the amplitudes, the integrated GPDs, to be extracted from the more easily measured imaginary parts. However, at non-zero momentum transfer DRs require integration over unphysical regions of the variables. We show that the relevant unphysical region of the non-forward DRs is considerable. This will vitiate the efforts to avoid the actual measurement of the real parts more directly. 
\end{abstract}

\maketitle


\section{Exclusive $\pi^0$ Electroproduction}
Data are accumulating on the cross section and asymmetries in exclusive $\pi^0$ electroproduction at intermediate energies $\nu$, low to high photon virtuality $Q^2$ and small momentum transfer $t$. This is the kinematical region in which both Regge and GPD descriptions overlap. Characteristic of this reaction is that it has chiral odd and C-parity odd t-channel quantum numbers. In the Regge picture the dominant exchanges will be neutral vector mesons ($\rho^0,\,\,\omega$ with $J^{PC}=1^{- -}$) and neutral axial vectors ($b_1^0,\,\,h_1$ with $1^{+ -}$) along with rescattering corrections. In the factorized GPD picture the chiral odd GPDs (soft physics) will be singled out for either pairs of quantum numbers and for both longitudinal and transverse photons (for the hard processes in the upper part of the handbag).
The quantum numbers of this reaction lead to the possibility of measuring the $transversity$ properties of the nucleon. 

A successful Regge cut model was developed to fit photoproduction data many years ago~\cite{GolOwe}. That model essentially involves as input the vector and axial vector meson trajectories that factorize into couplings to the on-shell $\gamma+\pi^0$ vertex and the nucleon vertex. The cuts or absorptive corrections destroy that factorization, but fill in the small $t$ and $t \approx -0.5$ amplitude zeroes. To connect to electroproduction, the upper vertex factor must acquire $Q^2$ dependence. This is accomplished by replacing the elementary, t-dependent couplings with $Q^2$ dependent transition form factors. In this Regge picture the factorization for the longitudinal virtual photon is not different from the transverse photon, except for the additional power of $Q^2$ for the longitudinal case. This is in contrast to the proofs of factorization for the longitudinal case in the GPD picture~\cite{CFS}. With our form factor approach to the upper vertex (including Sudakov factors to soften the endpoint singularities) we anticipate  a similar factorization for the transverse case. 

The Regge picture is implemented by singling out the 6 independent helicity amplitudes and noting that at large $s$ and small $|t|$ the leading natural parity and unnatural parity Regge poles contribute to opposite sums and differences of pairs of helicity amplitudes. This gives rise to striking behavior for certain observables.

Now the crucial connection to the 8 GPDs that enter the partonic  description of electroproduction is through the helicity decomposition~\cite{Diehl_01}, where, for example, one of the chiral even helicity amplitudes is given by 

$A_{++,++}(X,\xi,t)=\frac{\sqrt{1-\xi^2}}{2}(H^q+{\tilde H}^q-\frac{\xi^2}{1-\xi^2}(E^q+{\tilde E}^q)),$

while one of the chiral odd amplitudes is given by

$A_{++,--}(X,\xi,t)=\sqrt{1-\xi^2}(H_T^q+\frac{t_0-t}{4M^2}{\tilde H}_T^q-\frac{\xi}{1-\xi^2}(\xi E_T^q+{\tilde E}_T^q)).$

There are relations to PDFs, $H^q(X,0,0)=f_1^q(X)$, ${\tilde H}^q(X,0,0)=g_1^q(X)$, $H_T^q(X,0,0)=h_1^q(X)$. The first moments of these are the charge, the axial charge and the tensor charge, for each flavor $q$, respectively. Further, the first moments of $E(X,0,0)$ and $2{\tilde H}_T^q(X,0,0)+E_T^q(X,0,0)$ are the anomalous moments $\kappa^q, \kappa_T^q$, with the latter defined by Burkardt~\cite{Bur}.

\begin{figure}
  \includegraphics[height=.3\textheight]{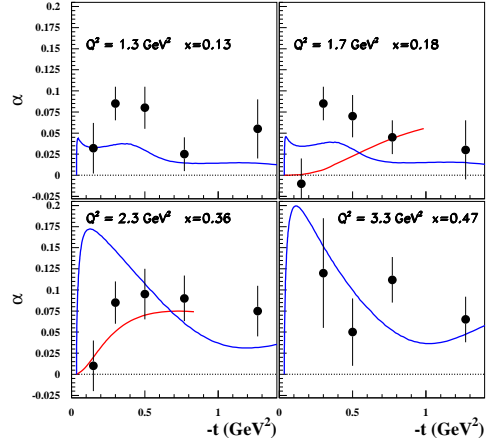}
  \caption{Beam asymmetry $\alpha$ in Regge (blue) and GPD pictures. Data are from Ref.~\cite{demasi}. }
\label{alpha}  
\end{figure}

\begin{figure}
  \includegraphics[height=.3\textheight]{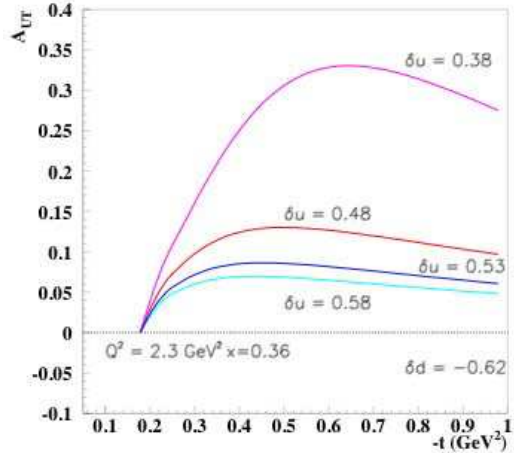}
  \caption{GPD predictions for transverse spin asymmetry, $A_{UT}$, for varying tensor charges. }
\label{aut}  
\end{figure}

Chiral even GPDs have been modeled in a thorough analysis~\cite{AHLT} , based on diquark spectators and Regge behavior at small $X$, and consistent with constraints from PDFs, form factors and lattice calculations. That analysis is used to obtain chiral odd GPDs via a multiplicative factor that fits the phenomenological $h_1(x)$~\cite{Anselmino}. With that {\it ansatz} the observables can be determined in parallel with the Regge predictions. In Fig.~\ref{alpha} the beam asymmetry is shown for both models and compared to recent data~\cite{demasi}. It is remarkable that these model predictions are not fit to that data, but determined by other constraints. Fig.~\ref{aut} shows a range of predictions from the chiral odd GPD model for the  target asymmetry, $A_{UT}$, depending on the choice of $\delta u$. This indicates that measurements will pin down values of the tensor charge. A thorough discussion of this work is given in Ref.~\cite{AGL}.

\section{GPDs and Dispersion Relations - Limitations}
 Deeply Virtual Compton Scattering (DVCS) amplitudes satisfy Lorentz covariance and analyticity in energy at fixed angles. It has long been known that analyticity of two body amplitudes leads to dispersion relations (DR). Using Mandelstam variables $s$  and $t$,  or $\nu = (s-u)/4M$ and $t$, we introduce a generic scattering amplitude $T(\nu,t,Q^2)$ to represent any of the invariant amplitudes in DVCS. For this discussion we can take all particles spinless. Then this amplitude would have its square magnitude proportional to $d\sigma/dt$ at fixed $Q^2,t$. The dispersion relation would be 
\begin{equation}
{\rm Re}T^-(\nu,t,Q^2)=\frac{2\nu}{\pi}\int_{\nu_{threshold}}^{\infty}d\nu^\prime\frac{{\rm Im}T^-(\nu^\prime,t,Q^2)}{{\nu^\prime}^2-\nu^2},
\label{nuDRodd}
\end{equation}
for an amplitude that is odd under the crossing symmetry $\nu\rightarrow -\nu$ (and thereby will avoid an extra subtraction for convergence) or
\begin{equation}
{\rm Re}T^+(\nu,t,Q^2)=\frac{\nu^2}{\pi}\int_{\nu_{threshold}}^{\infty}d\nu^\prime\frac{{\rm Im}T^+(\nu^\prime,t,Q^2)}{{\nu^\prime}^2({\nu^\prime}^2-\nu^2)} + \Delta,
\label{nuDReven}
\end{equation}
for an amplitude that is even under the crossing symmetry. The latter has a subtraction $\Delta$ to assure convergence in $\nu^\prime$. 

\begin{figure}
  \includegraphics[height=.3\textheight]{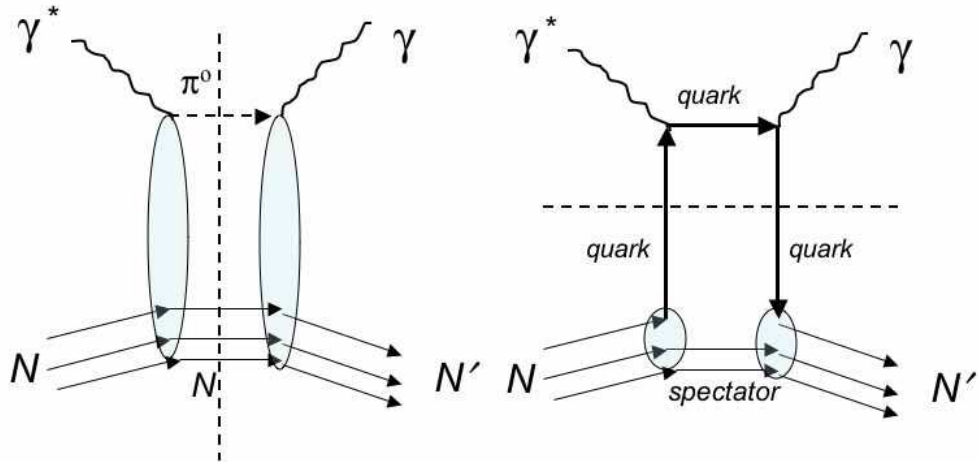}
  \caption{Left: hadronic s-channel unitarity cut for DVCS; Right: partonic ``handbag'' factorized diagram. }
\label{s-channel}  
\end{figure}

Where are the poles and branch cuts of $T^{\pm}$ as a function of $\nu^\prime$? Figure~\ref{s-channel} shows the lowest hadronic $continuum$ intermediate states, $\pi+N$. It is also clear that there will be a nucleon pole below the process threshold. The imaginary part of the amplitude is the discontinuity across the branch cut, beginning at $s=(M+m_\pi)^2$ and the corresponding value of $\nu=\nu_{continuum}$. As $\nu$ increases the next branch cut begins at the $2\pi+N$ threshold, and so on. There is a problem here, however. For non-forward values of $t$, the physically accessible threshold can begin at higher values of $\nu$ than $\nu_{continuum}$. But the DR requires an integrand defined down to the continuum. This was recognized to be a problem after non-forward DRs were first introduced and then studied carefully by Lehmann~\cite{Lehmann}. In the low energy region a partial wave decomposition allows for analytic continuation (via the ``Lehmann ellipse''). In the high energy region the partial wave approach is of no use, so a model of some kind must be used to complete the DR integration. We will show some examples below.

The important reason for considering these issues is that the possibility has been raised that DRs can be applied to the partonic, factorized form of the amplitudes~\cite{AniTer} - the handbag picture with GPDs of Fig.~\ref{s-channel}. An example is the spin independent  $H_f(x,\xi,t)$, for flavor $f$, which is related to the amplitude for DVCS~\cite{DieIva}
$T^{\mu\nu}(\nu,Q^2,t)$ through
\begin{equation}
T^{\mu\nu}(\nu,Q^2,t)=\frac{1}{2}g^{\mu\nu}{\bar u}(p^\prime){\hat n}u(p)
\sum_{flavors} e_f^2 {\cal H}_f (\xi,t),
\end{equation}
where 
\begin{equation}
{\cal H}_f (\xi,t)=\int_{-1}^{+1}dx \frac{H_f (x,\xi,t)}{x-\xi+i\epsilon},
\end{equation}
from which it follows (via $1/(x-\xi+i\epsilon)=P.V./(x-\xi)+\delta(x-\xi)/\pi$) that
\begin{equation}
\Im m({\cal H}_f(\xi,t))=H(\xi,\xi,t)\;\;
\mathrm{and}\;\;
\Re e({\cal H}_f(\xi,t)) =\frac{1}{\pi}P.V.\int_{-1}^{+1}dx  
\frac{H_f (x,\xi,t)}{x-\xi}.
\label{direct}
\end{equation}
With the relations among the independent variables
\begin{eqnarray}
x_{Bj}&= &\frac{Q^2}{2M\nu_{Lab}},\;\;  X=\frac{k\cdot P}{q\cdot P},\;\; \zeta=\frac{q\cdot P^\prime}{q\cdot P} \\
\xi&=&\frac{\zeta}{(2-\zeta)}=\frac{Q^2}{4M\nu},\;\; x=\frac{(2X-\zeta)}{(2-\zeta)} 
\label{kinematics}
\end{eqnarray} 
the DR in Eqn.~\ref{nuDReven} appear quite similar to Eqn.~\ref{direct};
\begin{equation}
\Re e({\cal H}_f(\xi,t)) =\frac{1}{\pi}P.V.\int_{-1}^{+1}dx  
\frac{H_f (x,\xi,t)}{x-\xi}=\frac{1}{\pi}P.V.\int_{-1}^{+1}dx  
\frac{H_f (x,x,t)}{x-\xi}+\Delta(\xi).
\label{DRgpd}
\end{equation}
It is this relation that led to the conclusion that using the Imaginary part of $\mathcal{H}$ is sufficient to determine the entire Compton Form Factor  $\mathcal{H}$. However, there are two possible problems. Consider Fig.~\ref{s-channel} where we see that the handbag factorized form of the amplitude does not have physical, on-shell observable intermediate states. It is not clear where the unitarity cut should begin in evaluating the DR for the GPD. Perhaps the relevant thresholds would be the jets associated with the partons, as suggested in Ref.~\cite{Collins}. But the threshold of such a branch cut is ill-defined. Putting that question aside for later exploration, we are left with the problem of the integration through an unphysical region between the continuum and the physical threshold for different values of $t$ and $Q^2$.

\begin{figure}
  \includegraphics[height=.35\textheight]{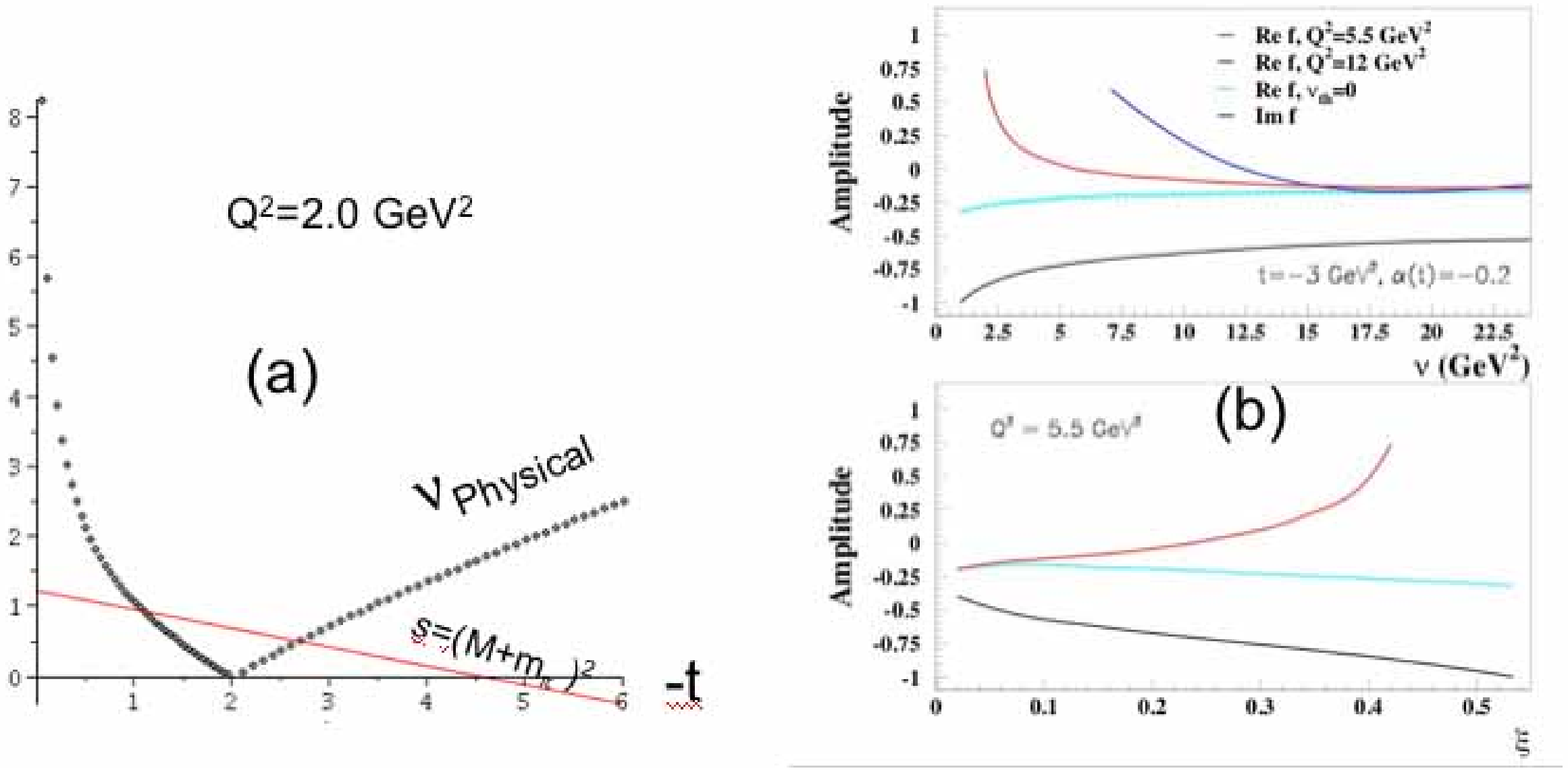}
  \caption{(a) Values of $\nu_{physical}$ vs. $-t$ for $Q^2=2$ GeV$^2$. The solid line is the DR lower limit for $s=(M+m_\pi)^2$. For $-1.2>t>-2.7$ GeV$^2$ there is no gap.(b) Dispersion relation for real part of  Regge amplitude for $\nu_{threshold}=0$ (third line down) which is exact, and for two other thresholds corresponding to different $Q^2$ values. The input imaginary part is the lower curve of both graphs.}
\label{nuth}  
\end{figure}

In Fig.~\ref{nuth} we show the mismatch between the physical threshold of $\nu$ for different $t$ values when, for exmple $Q^2=2$ GeV$^2$. This means that at $Q^2=2$ GeV$^2$ there are no measurements for $Im\mathcal{H}(\xi,t)$ for $\xi>\xi(\nu_{threshold})$. To complete the integration for the real part, though, requires continuing to $\xi=1$. How can this continuation be achieved? An example of the continuation for DVCS is a Regge pole model, continued down to $\nu=0$, 
\begin{equation}
H^R(\nu,Q^2,t)=\beta(Q^2,t)(1-e^{-i\pi\alpha(t)})(\nu/\nu_0)^{\alpha(t)},
\end{equation}
from which it follows that $Re H^R(\nu,Q^2,t)=tan(\pi\alpha(t)/2) Im H^R(\nu,Q^2,t)$ exactly. The DR relation is
\begin{equation}
ReH^R(\nu,Q^2,t)=\frac{2\nu}{\pi}\int_{\nu_{threshold}}^{\infty} d\nu^\prime \frac{Im H(\nu^\prime,Q^2,t)}{\nu^{\prime 2}-\nu^2}. 
\end{equation} 
The comparison of the actual real part with the result of the dispersion integration is shown in Fig.~\ref{nuth} for a few values of $Q^2$.
In terms of the variable $\xi$ it is clear that the dispersion integral only converges to the real part at very small $\xi$ or large $\nu$ (we have taken the odd amplitude here for simplicity).

We have checked this comparison with a simple, spinless diquark spectator model also. The model satisfies polynomiality and should thus satisfy the DR~\cite{AniTer}. Again we find that the convergence for non-zero $Q^2$ is slow because of the threshold mismatch. A more relevant model is the GPD parameterization of AHLT~\cite{AHLT}. We are in the process of checking the DR's for this determination of GPD's. In conclusion, DRs appear to be a parsimonious determination of the full analyticity of integrated GPDs. In practice, however, the unphysical region involved in implementing the integration requires model dependent interpolation.



\begin{theacknowledgments}
We thank the organizers of Spin 2008. For initial work we thank L. Gamberg and for many helpful comments and questions, C. Weiss,  J. Ralston, P. Stoler, H. Avakian, S. Ahmad. This work is supported by the U.S. Department of Energy grants no. DE-FG02-01ER4120 (S.L), and no. DE-FG02-92ER40702 (G.R.G.). \end{theacknowledgments}

\end{document}